\documentclass[final]{svjour2}
\usepackage{graphicx}
\usepackage{rotating}
\usepackage{amssymb}
\usepackage{mathptmx}
\usepackage{subfigure}
\usepackage[numbers]{natbib}
\usepackage{upgreek}

\makeatletter
\journalname{Journal of Low Temperature Physics}

\bibpunct{}{}{,}{s}{}{,}

\newcommand{\CMO}{CaMoO$_4$}

\begin{document}

\newcommand{\hdblarrow}{H\makebox[0.9ex][l]{$\downdownarrows$}-}
\title{Thermal Model and Optimization of a Large Crystal Detector using a Metallic Magnetic Calorimeter}

\author{G.B. Kim$^{1,2}$ \and S. Choi$^{2}$ \and Y.S.~Jang$^{1,{\dag}}$ \and H.J.~ Kim$^{3}$  \and Y.H.~Kim$^{1,4}$ \and V.V.~Kobychev$^{3}$ \and H.J.~ Lee$^{1,4}$ \and J.H.~Lee$^{1}$ \and J.Y.~ Lee$^{1,3}$ \and M.K.~Lee$^{1}$ \and S.J.~ Lee$^{1,{\ddag}}$ \and W.S. Yoon$^{1,4}$ }

\institute{1: Korea Research Institute of Standards and Science \\ 267 Gajeong-ro, Yuseong-gu, Daejeon 305-340, Republic of Korea.\\ Tel.:+82-42-868-5975\\
\email{yhkim@kriss.re.kr}\\
2: Seoul National University, Seoul 151-747, Republic of Korea\\
3: Kyungpook National University, Daegu 702-701, Republic of Korea\\
4: University of Science and Technology, Daejeon 305-333, Republic of Korea\\
$\dag$: Presently at Korea Atomic Energy Research Institute, Daejeon 305-353, Republic of Korea\\
${\ddag}$: Presently at NASA Goddard Space Flight Center, Greenbelt, MD 20771, USA\\
}

\date{\today}

\maketitle

\begin{abstract}
We established a simple thermal model of the heat flow in a large crystal detector designed for a neutrinoless double beta decay experiment. 
The detector is composed of a CaMoO$_{4}$ crystal and a metallic magnetic calorimeter (MMC). The thermal connection between the absorber and the sensor consists of a gold film evaporated on the crystal surface and gold bonding wires attached to this film and the MMC sensor. 
The model describes athermal and thermal processes of heat flow to the gold film.
A successive experiment based on optimization calculations of the area and thickness of the gold film showed a significant improvement in the size and rise-time of the measured signals.

\end{abstract}

\section{Introduction}

Over the last two decades, significant progress in the development of low temperature detectors for particle astrophysics applications has been made. Thermal detectors with dielectric crystal absorbers have been employed in many rare event experiments in deep underground laboratories around the world. These detectors can be designed to have high energy resolution, achieved by using sensitive temperature sensors such as neutron transmutation doped (NTD) Ge thermistors, transition edge sensors (TESs), and metallic magnetic calorimeters (MMCs)\cite{enss_book}. When semiconductor or scintillating crystals are used as the absorber, charge or light signals provide additional measurement channels along with temperature measurement (phonon channel). 
Simultaneous measurement of two different channels enables to reduce background events by different ionization or quenching factors for different particle types. 

Although any dielectric crystal can be used as an absorber of a thermal detector, the appropriate measurement of phonon signals is a non-trivial task. At low temperatures, the phonon transport through the interface between two solids becomes highly inefficient due to thermal boundary resistance. Moreover, if the temperature sensor measures the electron temperature, the energy of phonons must be transfered to the electron system. In the temperature sensor described above, weak electron-phonon interactions can be the major thermal impedance of phonon measurement at low temperatures. For example, the signal of an NTD Ge thermistor is slow to measure the instant temperature increase of a crystal. The rise-time can be extended as tens of milliseconds at a temperature of 10 mK.
The direct measurement of athermal phonons using TES sensors combined with superconducting phonon collectors on the crystal surface \cite{CDMS} may be an efficient way to overcome thermal impedances. However, the required fabrication processes for high performance TESs limit the choice of crystals.

We employed a MMC as the temperature sensor of a \CMO{} crystal. \CMO{} is a scintillating crystal that has the highest light yield among the studied molybdate crystals\cite{hjkim}. $^{100}$Mo within the crystal is a candidate for neutrinoless double beta decay, and its Q value is 3034.40(17)~keV\cite{Q_Mo100}.
Since the resolving power of a typical MMC used in x-ray measurement can be better than 2000, and the response time can be faster than 1~$\mu$s, MMCs can provide high performance in time and energy measurements in general\cite{ltd13_mmc}. MMCs can also present several advantages in detector development studies based on a given crystal. 
Many detector parameters can be adjusted to optimize the sensitivity under the experimental conditions during the design and operation stages. 
The present study aims to understand and manipulate the heat flow mechanism in \CMO{} thermal detectors suitable for neutrinoless double beta decay experiments\cite{sjlee}.

\section{Detector design and thermal model}
 \begin{figure}
        \centering
\subfigure[]
{
        \includegraphics[width=0.48\linewidth,keepaspectratio]{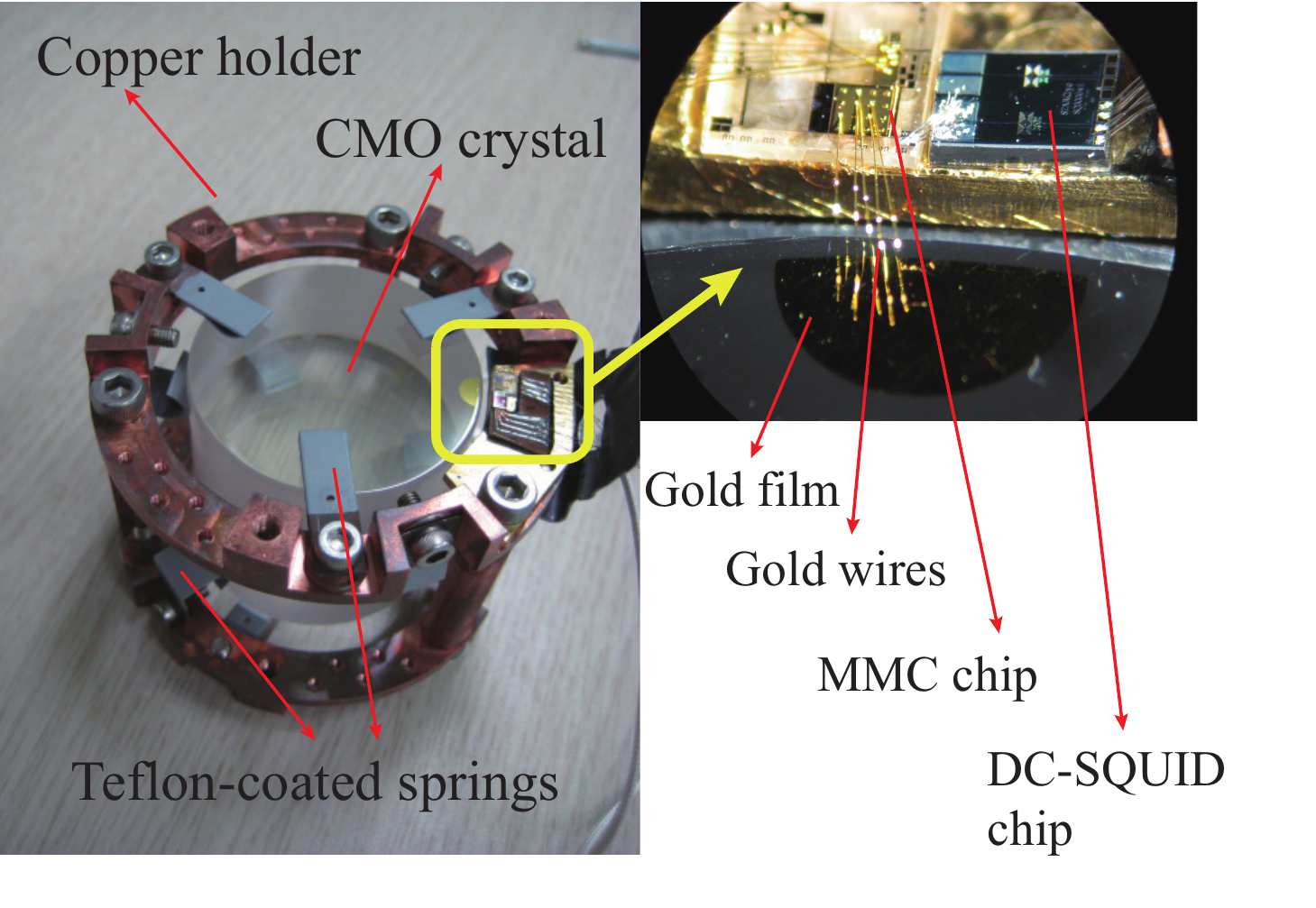}
        \label{fig:detector_design}
        }
        \subfigure[]
        {
                \includegraphics[width=0.48\linewidth,keepaspectratio]{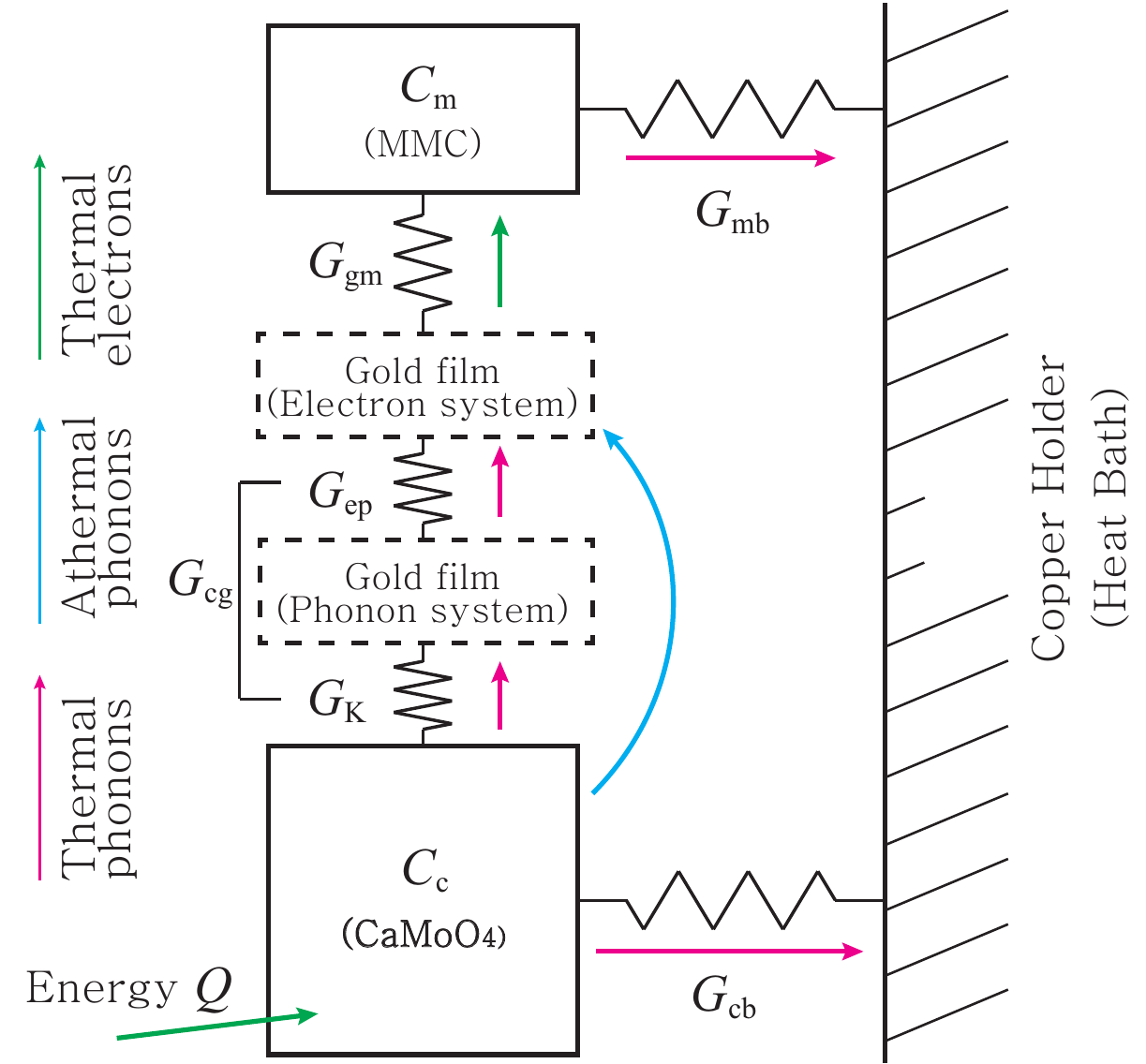}
        \label{fig:thermal_model}
        }
    \caption{(Color online) {\bf(a)} A picture of the first large {\CMO} crystal detector. The diameter of the crystal is 4 cm. {\bf(b)}A schematic diagram of the thermal model. }
    \label{fig:1}
\end{figure}
  A \CMO{} crystal with a diameter and height of 4 cm was used as an energy absorber. The crystal was held by Teflon-coated phosphor-bronze springs attached to a cylindrically structured copper sample holder, as shown in Fig.~\ref{fig:detector_design}. A micro-fabricated MMC chip\cite{Meander} produced in the Kirchhoff-Institute for Physics at Heidelberg University was employed as the temperature sensor. A gold film with an area and thickness of 10 mm$^2$ and 200 nm, respectively, was evaporated on the crystal surface close to the edge and was used for the thermal connection between the crystal and MMC sensor. Five 25~$\upmu$m gold wires were bonded to the gold film and Au:Er layer of the MMC chip to make a thermal connection. One advantage of the proposed design is that the heat flow between the absorber and the sensor can be controlled by changing the dimensions and location of the gold film and the number of gold bonding wires. Considerations and dimensions of the first designs were reported in detail elsewhere\cite{sjlee_thesis}.

We established a thermal model for the detector as shown in the schematic diagram illustrated in Fig.~\ref{fig:thermal_model}. The detector system is divided into three main components, the CaMoO$_{4}$ crystal, gold film, and MMC sensor. After particle absorption in the crystal, athermal phonons are initially generated in the crystal and travel around the crystal ballistically within a certain period of time\cite{yhkim}. 
A small portion of athermal phonons can be transmitted into the gold film. They can also be scattered by conduction electrons in the gold film and deposit their energy to conduction electrons.
The remaining athermal phonons are down-converted to a thermal phonon distribution over time. Subsequently, the energy transfer between the thermal phonons in the crystal and conduction electrons in the gold film follows thermal heat flow processes at low temperatures. Phonon transmission through the interface between \CMO{} and the gold film can be understood by acoustic mismatch model\cite{Kapitza}. Thermal phonons and electrons in the gold film interact with each other via electron-phonon interactions\cite{yhkim}. Heat flow along the gold wires between the gold film and MMC sensor, in which electrons in the Au:Er layer are strongly coupled to Er spins, occurs by electronic heat transfer according to the Weidemann-Franz law\cite{wsyoon}. 
$C_{\mathrm{c}}$, $C_{\mathrm{g}}$ and $C_{\mathrm{m}}$ are the heat capacities of the \CMO{}, the gold film, and the MMC, respectively. $G_{\mathrm{cg}}$, $G_{\mathrm{gm}}$, $G_{\mathrm{cb}}$ and $G_{\mathrm{mb}}$ are the thermal conductances between the crystal and gold film, the gold film and MMC, the crystal and heat bath, and the MMC and heat bath, respectively. Athermal and thermal processes in the thermal model are represented by long-curved and short-straight arrows in Fig.~\ref{fig:thermal_model}.

\begin{figure}
\centering
        \includegraphics[width=0.6\linewidth,keepaspectratio]{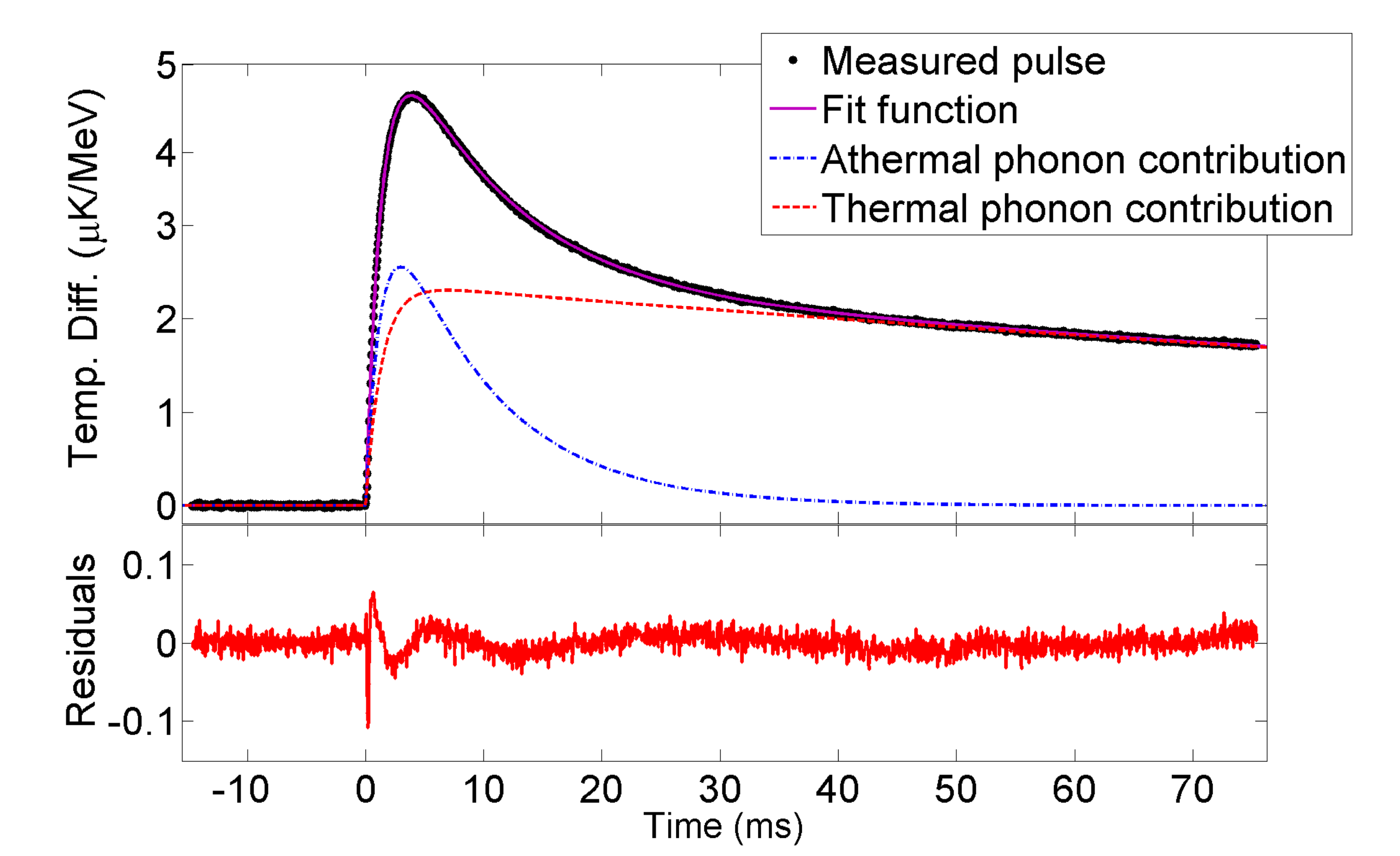}
    \caption{(Color online) Comparison of the measured pulse and the fitted result for 5.3 MeV surface alpha events. Athermal and thermal contributions of the signal are shown.}
    \label{fig:old_fit}
\end{figure}

The differential equations of heat flow in the thermal model can be written as:
\begin{eqnarray}
\label{thermal_eq}
\displaystyle (1-\epsilon){{Q} \over {\tau}}e^{-t/\tau} & = & C_{\mathrm{c}}{{dT_{\mathrm{c}}} \over {dt}} + (T_{\mathrm{c}}-T_{\mathrm{g}})G_{\mathrm{cg}}+(T_{\mathrm{c}}-T_{\mathrm{b}})G_{\mathrm{cb}}, \nonumber \\
\displaystyle \epsilon{{Q} \over {\tau}}e^{-t/\tau} & =& C_{\mathrm{g}}{{dT_{\mathrm{g}}} \over {dt}} + (T_{\mathrm{g}}-T_{\mathrm{c}})G_{\mathrm{cg}}+(T_{\mathrm{g}}-T_{\mathrm{m}})G_{\mathrm{gm}},\\
\displaystyle 0 & = & C_{\mathrm{m}}{{dT_{\mathrm{m}}} \over {dt}} + (T_{\mathrm{m}}-T_\mathrm{{g}})G_{\mathrm{gm}}+(T_{\mathrm{m}}-T_{\mathrm{b}})G_{\mathrm{mb}}, \nonumber
\end{eqnarray} 
where $Q$, $\epsilon$, $\tau$ and $T_{b}$ are the total input energy into the crystal, fraction of energy thermalized in the gold film by athermal phonon absorption, thermalization time of the crystal, and the temperature of the heat bath, respectively. 

The three simultaneous equations are equivalent to a third-order differential equation that has a complicated analytical solution. For simplicity, we assumed that the gold film did not have a heat capacity but only mediated heat flow. This is a valid approximation when $C_{\mathrm{g}} \ll C_{\mathrm{c}}, C_{\mathrm{m}}$.
Thus, Eq.~(\ref{thermal_eq}) can be reduced to second-order differential equations, and the solutions can be written as: 
\begin{eqnarray}
\displaystyle T_{\mathrm{c}} & = & A_{1} (e^{-t/a}-e^{-t/\tau}) + B_{1} (e^{-t/b} - e^{-t/\tau})+T_{\mathrm{b}}, \nonumber\\
\displaystyle T_{\mathrm{m}} & = & A_{2} (e^{-t/a}-e^{-t/\tau}) + B_{2} (e^{-t/b} - e^{-t/\tau})+T_{\mathrm{b}},
\end{eqnarray}
where $a$, $b$, $A_i$ and $B_i$ are functions of the thermal parameters. 
The temperature change of the MMC sensor, $(T_{\mathrm{m}} - T_{\mathrm{b}})$, can be considered as the sum of the two pulses that represent athermal and thermal contributions of the signal\cite{CRESST_thermal_model}. 
We attempted to fit $(T_{\mathrm{m}} - T_{\mathrm{b}})$ to the measured temperature signal by varying thermal parameters in the model. 
$C_{\mathrm{c}}$ and $C_{\mathrm{m}}$ were set to the calculated heat capacity of the \CMO{} crystal and MMC sensor, respectively, while $G$, $\epsilon$, and $\tau$ were used as fitting parameters in the analysis. The energy input $Q$ in Eq.~(\ref{thermal_eq}) corresponds to the phonon energy initiated by particle absorption.  We set $Q$ as 90\% of the total energy due to energy losses to scintillation lights, lattice damage, tunneling systems that their relaxation time is much longer than our signal decay-time, and unexpected thermal systems\cite{yhkim}. 

Fig.~\ref{fig:old_fit} shows the measured signal of 5.3~MeV alpha particles~\cite{sjlee_thesis} and the fitted results of a non-linear least-square fitting method. Both the rise and decay regions of the experimental and fitted signals were in good agreement. A similar analysis was repeated for measured pulses at 44 mK, 55 mK, and 60 mK. Consequently, $\epsilon$ and $\tau$ were approximately 2\% and 1.5~ms at 44 mK, respectively, and both $\epsilon$ and $\tau$ decreased slightly as the temperature increased. The values of $G$'s found in the analysis were in reasonable agreements and tendencies with expected values at different temperatures, and are not discussed further herein.

\section{Optimization of the detector design}
\begin{figure}
    \centering
    \subfigure[]
    {
        \includegraphics[width=0.48\linewidth,keepaspectratio]{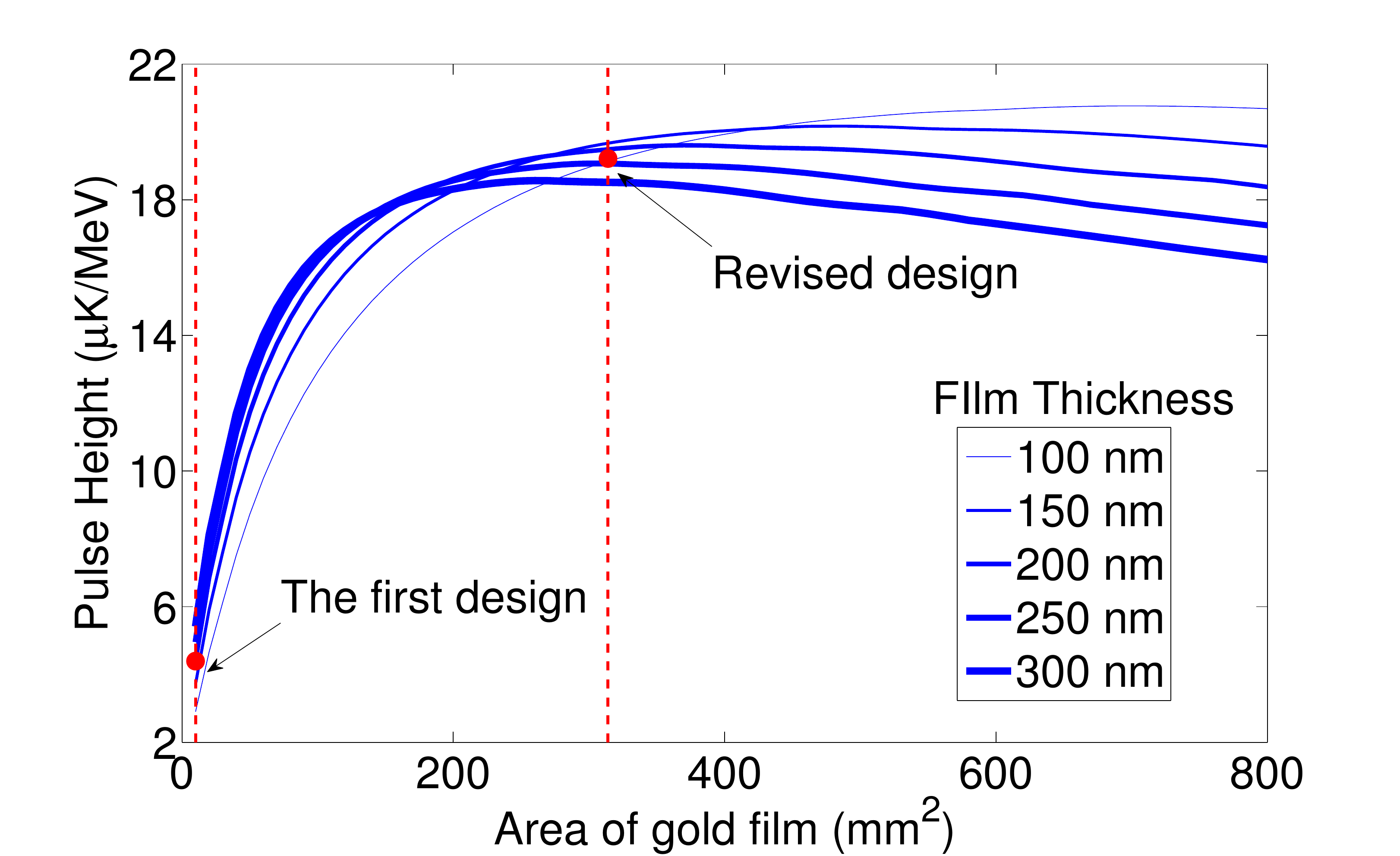}
        \label{fig:ph_goldfilm}
    }
    \subfigure[]
    {
        \includegraphics[width=0.48\linewidth,keepaspectratio]{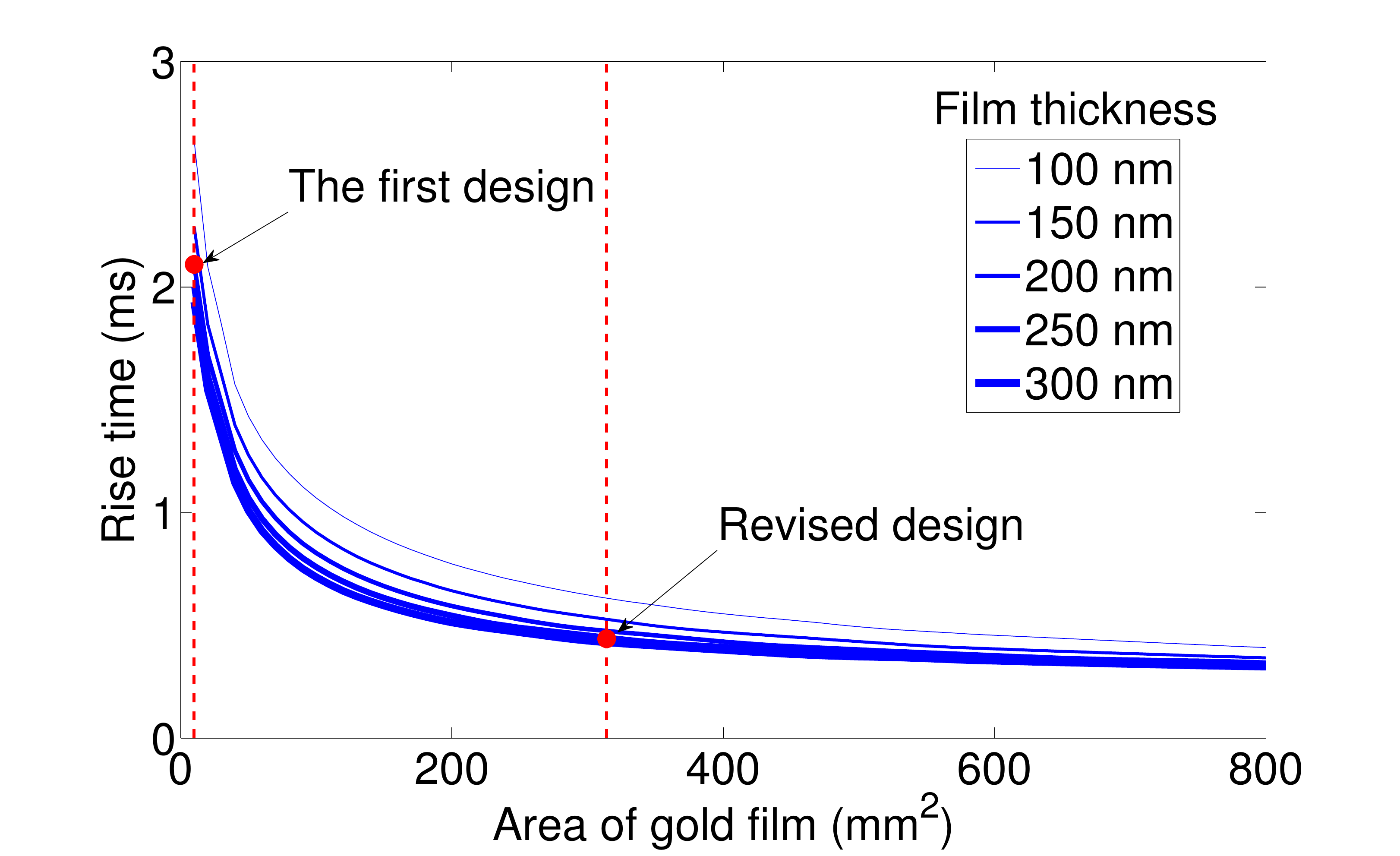}
        \label{fig:rise_goldfilm}
    }
    \caption{(Color online) {\bf{(a)}} Pulse height and {\bf(b)} rise-time of the calculated pulses for various gold film sizes at 44 mK.}
    \label{fig:optimization}
\end{figure}

The collection efficiency of athermal phonons depends on both the area and thickness of the gold film. Larger areas increase the chance that the initial athermal phonons hit the gold film. Thicker layers also increase the absorption probability of transmitted phonons to deposit their energy to electrons in the film.  
However, the signal size eventually becomes smaller with the larger heat capacity (or volume) of the film increased over a set of optimal dimensions. 
To investigate such dimensions of the gold film that maximize the signal size, the numerical solution of Eq.~(\ref{thermal_eq}) was calculated with different area and thicknesses of the gold film based on the parameters mentioned in the above analysis. Fig.~\ref{fig:optimization} shows the signal size and rise-time of the numerical solution for different film sizes. The time taken by a signal to change from 10\% to 90\% of the pulse height was considered the rise-time.

\begin{figure}
    \centering
    \subfigure[]
    {
        \includegraphics[width=0.47\linewidth,keepaspectratio]{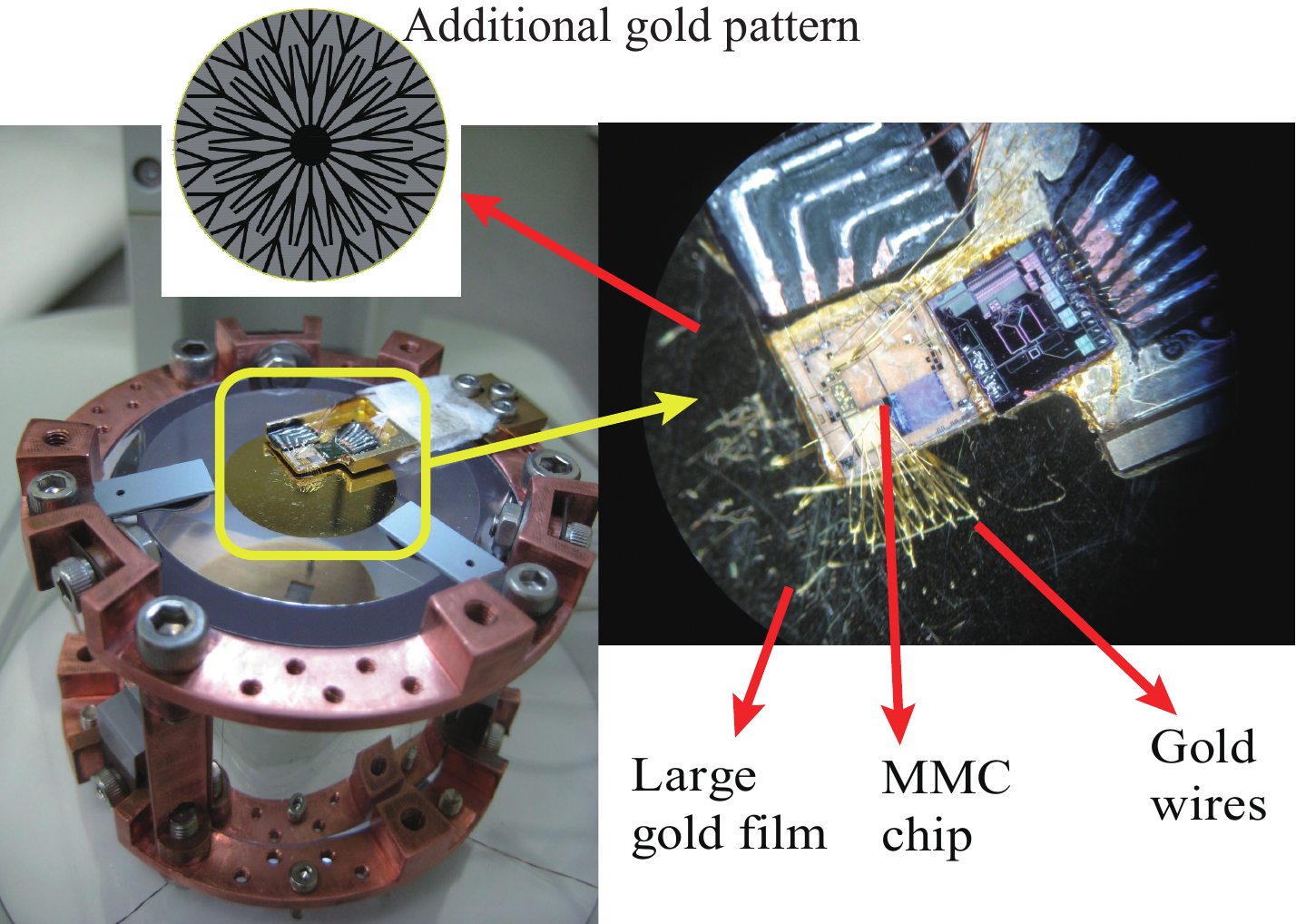}
        \label{fig:new_design}
    }
    \subfigure[]
    {
        \includegraphics[width=0.49\linewidth,keepaspectratio]{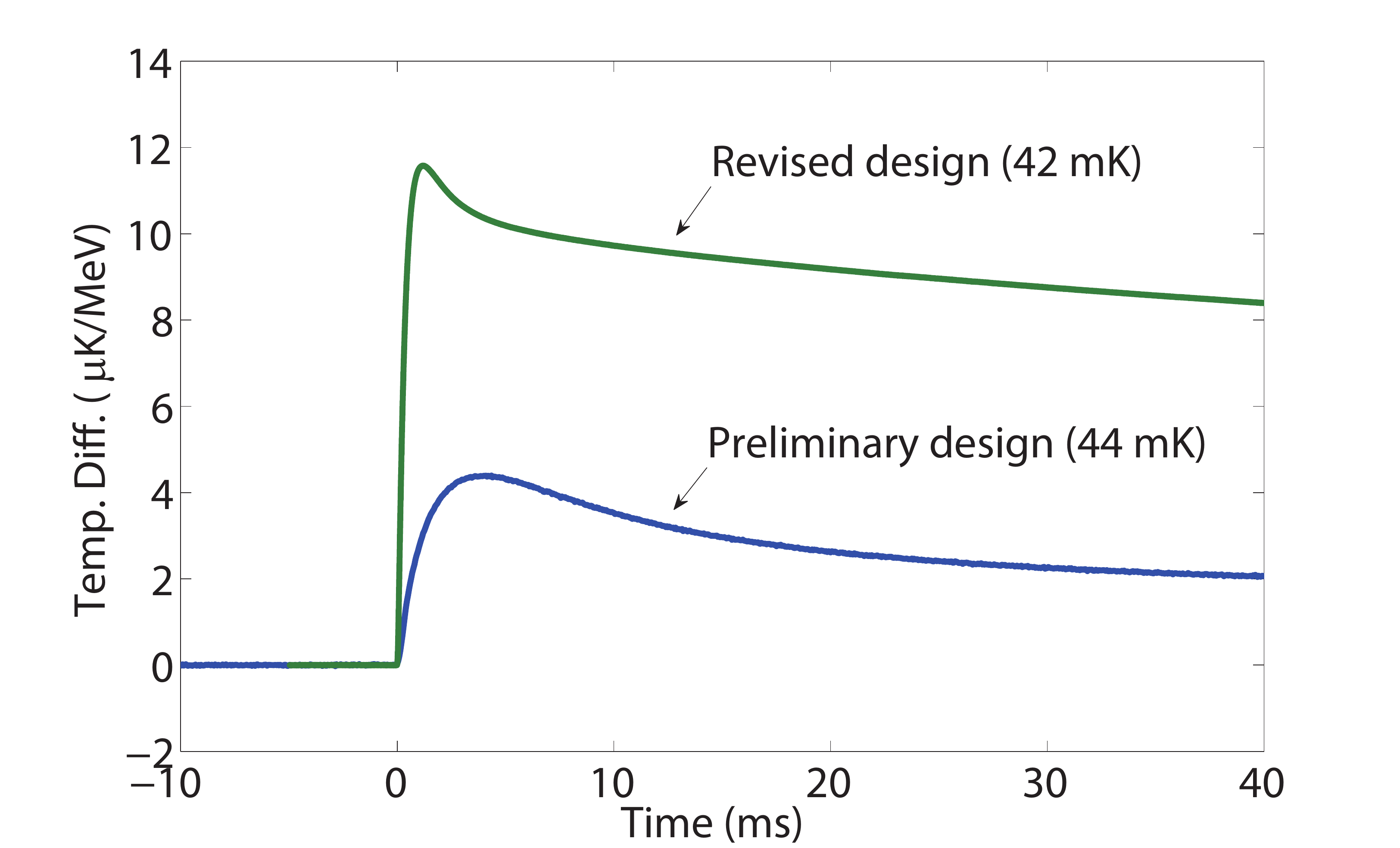}
        \label{fig:old_new}
    }
    \caption{(Color online) {\bf{(a)}} Revised design of the detector. Gray and black regions in the inserted figure indicate the large base gold film and an additional gold pattern, respectively. {\bf(b)} Comparison of pulse shapes obtained from the first and revised design. The operating temperature was 44 mk and 42 mK respectively.}
    \label{fig:5} 
\end{figure}

Based on the calculations of the expected pulses, we evaporated a new gold film in the middle of the surface of the \CMO{} crystal, as shown in Fig.~\ref{fig:new_design}. We chose a round film with a diameter and thickness of 2~cm and 200 nm, respectively. An additional pattern with the same thickness and 1/3 of the film area was evaporated to increase lateral thermal conductance of the gold film, and the expected value of $\epsilon$ was 39\%. 

The detector setup with the revised gold film showed larger and faster signals than the first design. The measured pulses in the two cases are compared in Fig.~\ref{fig:old_new}. The energy of signals were normalized to 1 MeV for comparison because the source used in the revised design had different energy with the first design.  
The normalized pulse height of the revised detector was about three times larger than that of the first design. The rise-time was 0.6~ms at 42~mK, which is close to the expected values shown in Fig.~\ref{fig:rise_goldfilm}.
Even though the heat capacity of the gold film increased by a factor of 31, the efficient collection of athermal phonons resulted in a larger and faster signal. 

The heat capacity of the revised design becomes comparable to that of the MMC. The approximation used for negligible heat capacity of the small gold film should not be applied in the thermal model of the revised detector. Further pulse shape studies and related details using the original thermal equations will be reported in future work.

The background spectrum measured with the revised setup showed an energy resolution of 9.9$\pm$1.3~keV FWHM for a full absorption line of 2.6 MeV($^{208}$Tl) gammas. Moreover, due to the rapid rise-time, alpha events could be discriminated from electron events using pulse shape analysis. Details on the measurement and analysis of the results will be discussed in a future report.

\section{Discussions and Conclusions}

We established a thermal model of a large crystal detector composed of a \CMO{} crystal and a MMC sensor for cryogenic particle detection. Heat flow to a gold film from the \CMO{} crystal was described by athermal and thermal processes. Based on the thermal model, an optimization simulation was performed for different sizes and thicknesses of the gold film. Significant improvements in the pulse size and rise-time were obtained in an experiment using an improved design.

However, the thermal model discussed herein assumes instant generation of athermal phonons after particle absorption in the crystal. Although the majority of absorbed energy was converted to phonons even in the scintillating crystals, some fraction of the phonons should be generated on a time scale related to the scintillation decay-time. A scintillation decay-time of about 0.4~ms was reported in a measurement with a \CMO{} crystal for alpha particles at 7~K\cite{mikhailik}. The lifetime of athermal phonons (or thermalization time, $\tau$) derived here may be partially affected by this generation of phonon. 
Moreover, the scintillation decay-time might be different for alpha and electron events\cite{camoo_scint}. The present analysis was performed for surface alpha signals; however, bulk electron events in \CMO{} may have different characteristics in phonon generation.

Although the current model does not cover all of the physical phenomena in the crystal, improvements were achived by understanding heat flow mechanisms in the detector. The detector performance with respect to energy and time resolutions in the present work demonstrated that cryogenic detection schemes of using \CMO{} and MMC are promising in the search for neutrinoless double beta decay. 

\section{Acknowledgements}
This work was supported by the National Research Foundation of Korea Grant funded by the Korean Government (NRF-2011-220-C00006).


\begin{thebibliography}{99}

\bibitem{enss_book}
C. Enss (Ed.) {\it Cryogenic Particle Detection}, Springer, (2005).

\bibitem{CDMS}
D. S. Akerib, {\it et al.} (CDMS Collaboration), {\it Phys. Rev. D} \textbf{72}, 052009, (2005).

\bibitem{hjkim} 
H.J. Kim, {\it et al.}, {\it IEEE T. Nucl. Sci.}  \textbf{57}, 1475, (2010). 



\bibitem{Q_Mo100}
S. Rahaman, {\it el al.}, {\it Phys. Lett. B} \textbf{662}, 111, (2008).

\bibitem{ltd13_mmc}
A. Fleischmann, {\it et al.},  {\it AIP Conf. Proc.} \textbf{571}, 1185, (1009).

\bibitem{sjlee} 
S.J. Lee, {\it et al.}, {\it Astropart. Phys.}  \textbf{34}, 732, (2011). 




\bibitem{Meander} A. Burck, {\it et al.}, {\it J. Low Temp. Phys.}  \textbf{151}, 337, (2008). 

\bibitem{sjlee_thesis} S.J. Lee, Ph.D. Dissertation, Seoul National University, (2012).

\bibitem{yhkim}
Y.H. Kim, {\it et al.},  {\it Nucl. Instrum. Meth. A} \textbf{520}, 208, (2004).

\bibitem{Kapitza}
E. T. Swartz, R. O. Pohl, {\it Rev. Mod. Phys.} \textbf{61}, 605, (1989).

\bibitem{wsyoon} Y.S. Yoon, {\it et al.}, {\it J. Low Temp. Phys.}  \textbf{167}, 280, (2012). 

\bibitem{CRESST_thermal_model}
F. Pr{\"o}bst, {\it el al.}, {\it J. Low Temp. Phys.}, \textbf{100}, 69, (1995).

\bibitem{mikhailik}
V.B. Mikhailik, {\it et al.}, {\it Nucl. Instrum. Meth. A} \textbf{583}, 350, (2007).

\bibitem{camoo_scint}
A.N. Annenkov, {\it et al.}, {\it Nucl. Instrum. Meth. A} \textbf{584}, 334, (2008).

\end{thebibliography}
\end{document}